\documentclass[12pt,notitlepage]{article}
\usepackage{epsfig, epsf, graphicx, subfigure}
\usepackage{amssymb,amsmath}
\usepackage{verbatim,enumerate}
\usepackage{rotating, lscape}
\usepackage{setspace}
\usepackage{bm}
\usepackage{pdflscape}
\usepackage[authoryear, sort]{natbib}

\def\log{\hbox{log}}

\def\boxit#1{\vbox{\hrule\hbox{\vrule\kern6pt
          \vbox{\kern6pt#1\kern6pt}\kern6pt\vrule}\hrule}}

\def\bse{\begin{eqnarray*}}
\def\ese{\end{eqnarray*}}
\def\be{\begin{eqnarray}}
\def\ee{\end{eqnarray}}
\def\bq{\begin{equation}}
\def\eq{\end{equation}}
\def\bse{\begin{eqnarray*}}
\def\ese{\end{eqnarray*}}

\newtheorem{Th}{Theorem}

\def\part{\partial}

\newcommand\eqdist{\mathrel{\stackrel{\makebox[0pt]{\mbox{\normalfont\scriptsize $d$}}}{=}}}
\begin{document}

\title{Density Deconvolution for Generalized Skew-Symmetric Distributions}
\author{Cornelis J. Potgieter \and Department of Statistical Science, Southern Methodist University, Dallas, TX}
\date{}
\maketitle

\begin{abstract}

This paper develops a density deconvolution estimator that assumes the density of interest is a member of the generalized skew-symmetric (GSS) family of distributions. Estimation occurs in two parts: a skewing function, as well as location and scale parameters must be estimated. A kernel method is proposed for estimating the skewing function. The mean integrated square error (MISE) of the resulting GSS deconvolution estimator is derived. Based on derivation of the MISE, two bandwidth estimation methods for estimating the skewing function are also proposed. A generalized method of moments (GMM) approach is developed for estimation of the location and scale parameters. The question of multiple solutions in applying the GMM is also considered, and two solution selection criteria are proposed. The GSS deconvolution estimator is further investigated in simulation studies and is compared to the nonparametric deconvolution estimator. For most simulation settings considered, the GSS estimator has performance superior to the nonparametric estimator.  	

\vspace{10mm}	
\end{abstract}

\pagenumbering{arabic} 


\section{Introduction}

The density deconvolution problem arises when it is of interest to estimate the density function $f_X(x)$ associated with a random variable $X$, but no $X$-sample is observed directly. Rather, the observed sample consists of contaminated data $W_j=X_j+U_j$, $j=1,\ldots,n$ where the $X_j$ are \textit{iid} with density $f_X$ and the $U_j$ are \textit{iid} random variables representing measurement error. This paper presents a semiparametric approach for estimating $f_{X}\left(x\right) $ that assumes the random variable $X$ belongs to the class of generalized skew-symmetric (GSS) distributions with a known symmetric component. The GSS deconvolution estimator explicitly models $X$ using as a base a symmetric parametric distribution and then uses kernel estimation methodology to estimate a skewing function which captures deviations from the base model. The GSS deconvolution estimator therefore attempts to capture the best of both parametric and nonparametric solutions.

The problem of estimating $f_X$ from a contaminated sample $W_1,\ldots,W_n$ was first considered by \cite{carroll1988optimal} and \cite{stefanski1990deconvolving}, who assumed that the distribution of the measurement error $U$ was fully known. Since then, much work on the topic has followed. \cite{fan1991asymptotic,fan1991optimal} considered the theoretical properties of the density deconvolution estimator and \cite{fan1993nonparametric} extended the methodology to nonparametric regression. \cite{diggle1993fourier} and \cite{neumann1997effect} considered the case of the measurement error distribution being unknown, and assumed that an external sample of error data was available to estimate the measurement error distribution. \cite{delaigle2008deconvolution} considered how replicate data can be used to estimate the characteristic function of the measurement error. The computation of the deconvolution density estimator also requires the selection of a bandwidth parameter. The two-stage plug-in approach of \cite{delaigle2002estimation} has become the gold-standard in application; \cite{delaigle2004practical} provides an overview of several popular bandwidth selection approaches. \cite{delaigle2008using} considered the use of simulation-extrapolation (SIMEX) for bandwidth selection in measurement error estimation problems.

Two recent papers considered the density deconvolution problem in new and novel ways. \cite{delaigle2014parametrically} considered parametrically-assisted nonparametric density deconvolution, while the groundbreaking work of \cite{delaigle2016methodology} made use of the empirical phase function to estimate the density function $f_X$ with the measurement error having unknown distribution and without the need for replicate data. The phase function approach imposes the restrictions that $X$ has no symmetric component and that the characteristic function of the measurement error is real-valued and non-negative.

The class of GSS distributions that forms the basis for estimatino in this paper has its roots in \cite{Azzalini1985}, which was the first publication discussing a so-called ``skew-normal'' distribution. There has been a great deal of activity since then with the monographs by \cite{genton2004skew} and \cite{Azzalini2013} providing a good overview of the existing literature on the topic. Much of the GSS research has been theoretical in nature and while such theoretical work is important for understanding the statistical properties of GSS distributions, the applied value associated with this family of distributions has not often been realized in the literature. Notable and relevant exceptions that have used GSS distributions in application have considered the modeling of pharmacokinetic data, see \cite{chu2001stat}, the redistribution of soil in tillage, see \cite{van2003simulation}, and the retrospective analysis of case-control studies, see \cite{guolo2008flexible}.  All of these authors considered only fully parametric models and therefore did not exploit the flexibility of GSS distributions as a semiparametric modeling tool. \cite{arellano2005skew} considered a fully parametric measurement error model assuming both the distribution of $X$ and $U$ follow skew-normal distributions. \cite{lachos2010multivariate} modeled $X$ using a scale-mixture of skew-normal distributions while assuming $U$ follows a mixture of normals distribution. No other work applying GSS distributions in the measurement error context was found.


\section{Generalized Skew-Symmetric Deconvolution Estimator}

Consider the problem of estimating the density function $f_X(x)$ associated with random variables $X$ based on a sample contaminated by additive measurement error $W_j=X_j+U_j$, $j=1,\ldots,n$. The random variables $W$ and $U$ represent, respectively, the contaminated observation and the measurement error. Let $f_{U}\left( u\right) $ denote the density function of the measurement error $U$, which is assumed to have a symmetric distribution with mean $\mathrm{E}(U)=0$ and variance $\mathrm{Var}(U)=\sigma_U ^{2}$. As is standard in deconvolution problems, the distribution of $U$ is assumed known. If replicate observations were available, one would be able to estimate the distribution of $U$ and/or the value of $\sigma_U^2$. Estimation of the measurement error distribution is well-established in the literature, see for example \cite{delaigle2008deconvolution}, and is tangential to the development of the skew-symmetric deconvolution estimator presented here.

Next, assume that the random variable of interest $X$ can be expressed as $X=\xi +\omega Z$, where $\xi \in \mathbb{R} $ and $\omega >0$ are, respectively, location and scale parameters and the random variable $Z$ has density function%
\begin{equation}
f_{Z}\left( z\right) =2f_{0}\left( z\right) \pi \left( z\right) \text{, }%
z\in \mathbb{R}  \label{GSS density}
\end{equation}%
where $f_{0}\left( z\right) $ is a density function symmetric around $0$ and $\pi
\left( z\right) $, hereafter referred to as the skewing function, satisfies the constraints%
\begin{equation}
0\leq \pi \left( z\right) =1-\pi \left( -z\right) \leq 1.
\label{Constraint on pi}
\end{equation}%
The approach considered in this paper is semiparametric in that the symmetric pdf $f_{0}\left(
z\right) $ is assumed known, but no assumptions regarding the skewing
function $\pi \left( z\right) $ are made beyond adherence to the constraint \eqref{Constraint on pi}. Any function satisfying this constraint can be paired with a symmetric pdf $f_0(z)$ and will result in \eqref{GSS density} being a valid pdf. The GSS representation provides a very flexible model for density $f_X(x)$. Specifically, Appendix A.1 shows that for each real value $\xi$, there is a triple $(\omega_\xi,f_\xi(z),\pi_\xi(z))$ with $f_\xi(z)$ symmetric about 0, and $\pi_\xi(z)$ satisfying \eqref{Constraint on pi} such that $X=\xi+\omega_\xi Z_\xi$ with $Z_\xi$ having pdf $2f_\xi(z)\pi_\xi(z)$. This holds for any continuous random variable $X$. As such, $X$ has an infinite number of skew-symmetric representations and it is necessary to specify the symmetric component $f_0(z)$ for identifiability. If this specification were not made, only two of the three model components $\left(\xi,\omega\right)$, $f_0(z)$ and $\pi(z)$ would be uniquely identifiable even with known pdf $f_X(x)$.

An important property of GSS random variables that plays a central role in the development of an estimator is an invariance under even transformations. Let $Z$ be a GSS random variable according to (\ref{GSS density}) and let $Z_0$ have symmetric density $f_0$. For any even function $t(z)$, it holds that
\[t(Z)\eqdist t(Z_0),\] where $\eqdist$ denotes equality in distribution, see for example Proposition 1.4 in \cite{Azzalini2013}. That is, no knowledge of the skewing function $\pi(z)$ is required to determine the distribution of $t(Z)$ with known symmetric component $f_0(z)$.

Now, let $\psi _{Z}\left( t\right) $ denote the characteristic function of the
random variable $Z$, and let \[
c_{0}\left( t\right) = \mathrm{Re}[\psi_Z(t)], \quad s_{0}\left( t\right) = \mathrm{Im}[\psi_Z(t)]\] denote the real and
imaginary components of the characteristic function. The real component can be expressed as $c_{0}\left(
t\right) = \mathrm{E}\left[\cos(tZ)\right]$. By the property of even transformation of $Z$, $c_{0}\left( t\right) $ is the characteristic function of the symmetric
density $f_{0}\left( z\right) $. 

The development of the GSS deconvolution estimator in the remainder of this section, as well as bandwidth estimation in Section \ref{sec:bandwidth}, will proceed assuming the pair $\left(\xi,\omega\right) $ is known. Estimation of these location and scale constants will be discussed in Section \ref{sec:parm estim}. 

Let $W^{\ast }=\omega ^{-1}\left( W-\xi \right)$ denote the translated and rescaled $W$ and observe that $W^{\ast } = Z+\omega^{-1}U$. The random variable $W^{\ast }$ therefore has characteristic
function $\psi _{W^{\ast }}\left( t\right) =\psi _{Z}\left( t\right) \psi
_{U}\left( t/\omega \right) $, where $\psi _{U}\left( t\right) $ is the real-valued characteristic
function of the measurement error $U$. It follows that
\begin{equation}
\mathrm{Re}\left\{ \psi _{W^{\ast }}\left( t\right) \right\} =c_{0}\left(
t\right) \psi _{U}\left( t/\omega \right)  \label{real w_star}
\end{equation}%
and%
\begin{equation}
\mathrm{Im}\left\{ \psi _{W^{\ast }}\left( t\right) \right\} =s_{0}\left(
t\right) \psi _{U}\left( t/\omega \right)  \label{imaginary w_star}
\end{equation}%
where the functions $c_0(t)$ and $\psi_U(t)$ in (\ref{real w_star}) are known, but the function $%
s_{0}\left( t\right) $ in (\ref{imaginary w_star}) is unknown. If an appropriate
estimator of $s_{0}\left( t\right) $ can be obtained, the density function of $Z$, and consequently the density function of $X$, can also be estimated.

In considering estimation of $s_{0}\left( t\right) $, note that the pdf $f_Z(z)$ can be expressed as
\begin{eqnarray}
f_{Z}\left( z\right) &=&\frac{1}{2\pi }\int_{\mathbb R }\exp \left(
-itz\right) \psi_{Z}\left( t\right) dt \notag  \\
&& \notag \\
&=&\frac{1}{2\pi }\int_{\mathbb R }\exp \left( -itz\right) \left\{ c_{0}\left(
t\right) +is_{0}\left( t\right) \right\} dt \notag \\
&& \notag  \\
&=&f_{0}\left( z\right) +\frac{1}{2\pi }\int_{\mathbb R }\sin \left( tz\right) 
s_{0}\left( t\right) dt, \label{fZ_ito_s0}
\end{eqnarray}
where  the second equality simply expresses the characteristic function in terms of its real and imaginary components, while the last equality follows from some straightforward algebra. The importance of equation (\ref{fZ_ito_s0}) follows on noting that any estimator of $s_0(t)$, say $\check{s}_0(t)$, needs to have a well-defined (finite) integral $\int \sin (tz) \check{s}_0(t) dt$ for all $t$ in order for an estimator of $f_Z(z)$ to be valid.

For random sample $W_1,\ldots,W_n$, let $W_j^\ast = (W_j-\xi)/\omega$ for $j=1,\ldots,n$. Consider the standard empirical estimator of $s_0(t)$,%
\begin{equation*}
\tilde{s}_{0}\left( t\right) =\frac{1}{\psi _{U}\left( t/\omega\right) }%
\frac{1}{n}\sum_{1\leq j\leq n}\sin \left( tW_{j}^{\ast }\right) .
\label{emp_s0}
\end{equation*}%
This empirical estimator, while unbiased for $s_0(t)$, does not have well-defined integral when substituted in (\ref{fZ_ito_s0}). This can be understood by the looking at the tail behavior of $\tilde{s}_{0}\left( t\right) $. For any continuous distribution, the true function $s_0(t)$ will eventually converge to $0$ as $\left\vert t\right\vert \rightarrow \infty $. On the other hand, $\sum_j \sin (tW_j^{\ast})$ is periodic and after division by ${\psi _{U}\left( t/\omega\right) }$,  $\tilde{s}_0(t)$ blows up as $|t|$ increases. 

As an alternative to the unbiased empirical estimator, consider the a smoothed estimate, 
\begin{equation}
\hat{s}_{0}\left( t\right) =\frac{\psi _{K}\left( ht\right) }{\psi
_{U}\left( t/\omega\right) }\frac{1}{n}\sum_{1\leq j\leq n}\sin \left(
tW_{j}^{\ast }\right)  \label{smooth s0}
\end{equation}%
where $\psi _{K}\left( t\right) $ is a non-negative weights function and $h$ is a bandwidth parameter. This smoothed estimator is biased for $s_0(t)$, $\mathrm{E}[\hat{s}_{0}(t)]=\psi_K(ht)s_0(t)$, but has several other properties that are desirable. Firstly, it is an odd function, $\hat{s}_0(-t)=-\hat{s}_0(t)$ for all $t \in \mathbb R$. Secondly, substitution of (\ref{smooth s0}) into (\ref{fZ_ito_s0}) results in well-defined estimator provided $\psi_K(t)$ is chosen with some care. It is necessary to have $\left\vert \psi _{K}\left(
ht\right) /\psi _{U}\left( t/\omega \right) \right\vert \rightarrow 0$ as $%
\left\vert t\right\vert \rightarrow \infty $. Any function $\psi_K(t)$ that is equal to 0 outside a bounded interval will trivially satisfy this requirement. Thus, an estimate of the density $f_Z(z)$ based on \eqref{smooth s0} is given by
\begin{equation}
\hat{f}_Z(z) = f_0(z) + \frac{1}{2\pi}\int_{\mathbb R} \sin(tz)\hat{s}_0(t)dt. \label{fZ_est1}
\end{equation}

Estimator \eqref{fZ_est1} suffers from the same drawback as the usual nonparametric deconvolution estimator in that it may be negative in parts. Therefore, when estimator \eqref{fZ_est1} is used, the negative parts of the estimated density should be truncated and the positive part of the function rescaled to integrate to $1$. Additionally, the integral form of \eqref{fZ_est1} is not computationally convenient. Note, however, that by combining equations (\ref{GSS density}) and (\ref{fZ_ito_s0}), it is possible to write the skewing function as
\begin{equation}
\pi \left( z\right) =\frac{1}{2}-\frac{1}{4\pi f_{0}\left( z\right) }%
\int_{R}\sin \left( tz\right) s_{0}\left( t\right) dt. \label{pi_ito_s0}
\end{equation}%
Substitution of the smoothed estimator (\ref{smooth s0}) in (\ref{pi_ito_s0}) and recalling that $\sin(tz)=\left(e^{itz}-e^{-itz} \right)/(2i)$, it is easily verified that the resulting estimate of the skewing function is given by
\begin{equation}
\hat{\pi}\left( z\right) =\frac{1}{2}+\frac{1}{8f_{0}\left( z\right) }%
\left\{ \tilde{f}_{W^*}\left( z\right) -\tilde{f}_{W^*}\left( -z\right) \right\} \label{pi_est1}
\end{equation}
where
\begin{equation*}
\tilde{f}_{W^*}\left( z\right) =\frac{1}{nh\omega}\sum K_{h\omega}\left( \frac{%
z-W_{j}^*}{h\omega}\right)
\end{equation*}
is the nonparametric deconvolution density estimator of \cite{carroll1988optimal} and \cite{stefanski1990deconvolving} with deconvolution kernel 
\begin{equation*}
K_{h}\left( y\right) =\frac{1}{2\pi }\int_{\mathbb R } e^{-ity}\frac{\psi _{K}\left( t\right) }{\psi _{U}\left( t/h\right) 
}dt.
\end{equation*}%

While the estimator $\hat{%
\pi}\left( z\right) $ satisfies the required relationship $\hat{\pi}%
\left( -z\right) =1-\hat{\pi}\left( z\right) $, it is not range-respecting.
Specifically, it is possible to have $\hat{\pi}\left( z\right)
\not\in \left[ 0,1\right] $ for a set $z$ with nonzero measure. A finite-sample correction needs to be applied. It is recommended that the range-corrected skewing function
\begin{equation}
\tilde{\pi}\left( z\right) =\max \left\{ 0,\min \left\{ 1,\hat{\pi}\left(
z\right) \right\} \right\} \label{pi_est2}
\end{equation}%
be used to estimate $\pi(z)$, with the corresponding estimate of the density function of $Z$ being%
\begin{equation*}
\tilde{f}\left( z\right) =2f_{0}\left( z\right) \tilde{\pi}\left( z\right) .
\end{equation*}%
The estimated density function of $X$ is therefore%
\begin{equation}
\tilde{f}\left( x|\xi ,\omega \right) =\frac{1}{\omega }f_{0}\left( \frac{%
x-\xi }{\omega }\right) \tilde{\pi}\left( \frac{x-\xi }{\omega }\right) .
\label{f_tilde_X}
\end{equation}%
Use of the corrected skewing function estimate \eqref{pi_est2} ensures that \eqref{f_tilde_X} is always a valid density function. There is no need for any additional truncation of negative values and subsequent rescaling as would be the case with direct implementation of \eqref{fZ_est1} or with the usual nonparametric deconvolution estimator.

\section{Some Properties of the GSS Deconvolution Estimator}

The estimator $\hat{\pi}(z)$ in (\ref{pi_est1}) is a consistent estimator of $\pi(z)$ for appropriately chosen bandwidth $h$. Specifically, using known properties of the nonparametric deconvolution estimator $\tilde{f}_{W^\ast}(z)$ and equation \eqref{pi_est1}, it follows that
\begin{equation*}
E\left[ \hat{\pi}\left( z\right) \right] =\pi \left( z\right) -\frac{c_{K}%
	\left[ f_{Z}^{\prime \prime }\left( -z\right) -f_{Z}^{\prime \prime }\left(
	z\right) \right] }{4f_{0}\left( z\right) }h^{2}+O\left( h^{3}\right)
\end{equation*}%
with $c_{K}$ a constant only depending on the kernel function $\psi_K(t)$. The adjusted estimator $\tilde{\pi}(z)$ in (\ref{pi_est2}) is asymptotically equivalent to (\ref{pi_est1}), and therefore the density estimator $\tilde{f}(x|\xi,\omega)$ in (\ref{f_tilde_X}) is asymptotically unbiased for $f_X(x)$ for appropriate bandwidth $h$.

Central to further understanding of the properties of the GSS deconvolution density estimator is the smoothed estimate of the imaginary component of the characteristic function, $\hat{s}_0$. As stated in the previous section, this estimator has expectation $\mathrm{E}\left[\hat{s}_0(t)\right] = \psi_K(ht)s_0(t)$. Additionally, it has covariance structure
\begin{multline} \label{Cov_s0}
\mathrm{Cov}\left[\hat{s}_0(t_1),\hat{s}_0(t_2)\right]  =  \frac{\psi_K(ht_1)\psi_K(ht_2)}{n} \\
\times \left[\frac{c_0(t_1-t_2)\psi_U((t_1-t_2)/\omega)-c_0(t_1+t_2)\psi_U((t_1+t_2)/\omega)}{2\psi_U(t_1/\omega)\psi_U(t_2/\omega)}-s_0(t_1)s_0(t_2)\right]
\end{multline}
The integrated squared error (ISE) of the GSS estimator can also be expressed in terms of $\hat{s}_0(t)$,
\begin{equation*}
\begin{split}
\mathrm{ISE} &  = \int_{\mathbb{R}} \left[\tilde{f}_Z(z)-f_Z(z)\right]^2dz \\
& = \frac{1}{2\pi} \int_{\mathbb{R}} \left|c_0(t)+i\hat{s}_0(t)-\psi_Z(t)\right|^2dt \\
& = \frac{1}{2\pi} \int_{\mathbb{R}} \left[\hat{s}_0(t)-s_0(t)\right]^2dt
\end{split}
\end{equation*} 
where the first equality follows from application of Parseval's identity and the second upon noting that the real component $c_0(t)$ is common to the estimated and true characteristic functions. The mean integrated square error, $\mathrm{MISE}=\mathrm{E}[\mathrm{ISE}]$, is a function of the bandwidth $h$, and using $\mathrm{E}\left[\hat{s}_0(t)\right]$ and (\ref{Cov_s0}), the latter upon setting $t_1=t_2=t$, it follows that
\begin{equation} \label{MISE}
\mathrm{MISE}(h) = \left(2\pi\right)^{-1}\int_{\mathbb{R}} \left\{\frac{\psi_K^2(ht)}{n} \left[\frac{1-c_0(2t)\psi_U(2t/\omega)}{2\psi_U^2(t/\omega)}-s_0^2(t)\right]+\left[\psi_K(ht)-1\right]^2s_0^2(t)\right\} dt.
\end{equation}
A special distributional case that is of particular interest is the symmetric one. In this instance, $s_0(t)=0$ for all $t$. The $\mathrm{MISE}$ in (\ref{MISE}) then becomes
\begin{align*}
\mathrm{MISE}_{\mathrm{sym}}(h) & = \left(4\pi\right)^{-1} \int_{\mathbb{R}} \frac{\psi_K^2(ht)}{n} \left[\frac{1-c_0(2t)\psi_U(2t/\omega)}{\psi_U^2(t/\omega)}\right] dt \\
& \leq \left(2\pi n\right)^{-1} \int_{\mathbb{R}} \frac{\psi_K^2(ht)}{\psi_U^2(t/\omega)}dt.
\end{align*}
where the inequality follows upon noting that $\vert {1-c_0(2t)\psi_U(2t/\omega)} \vert \leq 2$ for all $t$.  The upper bound of $\mathrm{MISE}_{\mathrm{sym}}$ is proportional to the asymptotic $\mathrm{MISE}$ of the nonparametric deconvolution estimator, see for example equation (2.7) in Stefanski \& Carroll (1990). This suggests that, in the symmetric case, one could expect the GSS deconvolution estimator to perform better than the nonparametric deconvolution estimator if the symmetric component $c_0(t)$ has been correctly specified. The $\mathrm{MISE}$ in (\ref{MISE}) will be revisited in Section \ref{sec:bandwidth} when considering estimating of the bandwidth for GSS deconvolution.

\section{Bandwidth Selection} \label{sec:bandwidth}

Two bandwidth selection approaches will be developed in this section. The first is a cross-validation approximation to the ISE, while the second is a method for approximating the $\mathrm{MISE}$ in (\ref{MISE}).

\subsection{Cross-Validation Bandwidth}

Recall that by Parseval's identity,
\begin{equation}
\int_{\mathbb R} \left[\tilde{f}_0(z)-f_0(z)\right]^2dz \propto \int_{\mathbb R} \left[\hat{s}_0(t)-s_0(t)\right]^2 dt. \label{parseval_prop}
\end{equation}
where $\propto$ indicates proportionality. Let $C(h)$ be the expression obtained by expanding the square on the right-hand side of \eqref{parseval_prop} and keeping only terms involving the estimator $\hat{s}_0(t)$; that is,
\begin{equation} \label{C_h}
	C(h) = \int_{\mathbb R} \hat{s}_0^2(t)dt - 2 \int_{\mathbb R} \hat{s}_0(t) s_0(t)dt.
\end{equation}
Now, note that the second integral in \eqref{C_h} can be written as
\begin{equation}
\int_{\mathbb R} \hat{s}_0(t) s_0(t)dt = \sum_{i=1}^n \int_{\mathbb R} \frac{\psi_K(ht)\sin(tW_i^\ast)}{\psi_U(t/\omega)} s_0(t)dt. \label{s0 integral expand}
\end{equation}
Define
\[\tilde{s}_{(i)}(t) = \frac{(n-1)^{-1}\sum_{j\neq i} \sin (tW_j^*)}{\psi_U(t/\omega)}, \] the empirical estimate of $s_0(t)$ excluding the $i$th observation. The quantity $\tilde{s}_{(i)}(t)$
is an unbiased estimator of $s_0(t)$ independent of $W_i$. The cross-validation score follows by substitution of $\tilde{s}_{(i)}(t)$ in \eqref{s0 integral expand} for each $i$ in the summation, and subsequently an estimate of \eqref{C_h} is
\begin{equation} \label{CV_score}
\hat{C}(h) = \int_{\mathbb R} \frac{\psi_K(ht)}{\psi_U^2(t/\omega)}\left[\psi_K(ht)\left(\frac{1}{n}\sum_{j=1}^{n}\sin (tW_j^*)\right)^2-\frac{2}{n(n-1)}\sum_{i=1}^{n}\sum_{j\neq i}\sin(tW_i^*)\sin(tW_j^*)\right].
\end{equation}
The CV score \eqref{CV_score} is similar to that of \cite{stefanski1990deconvolving} in the nonparametic setting, except it only depends on estimating the imaginary component of the characteristic function. The CV bandwidth estimate is the value $\tilde{h}$ that minimizes $\hat{C}(h)$.

\subsection{Approximate MISE Bandwidth}

The second bandwidth approach considered is one that finds an estimator of the MISE in (\ref{MISE}) that can be minimized. The only unknown quantity in \eqref{MISE} is $s_0^2(t)$. Note that \[\mathrm{E}\left[\sin(tW_j^*)\sin(tW_k^*)\right]=\psi_U^2(t/\omega)s_0^2(t)\] whenever $j \neq k$. Thus, the square of the imaginary component, $s_0^2(t)$, can be estimated by
\begin{equation} \label{s2_est}
\hat{s_2}(t) = \max \left\{0, \frac{1}{n(n-1)\psi_U^2(t/\omega)}\sum_{j=1}^{n} \sum_{k \neq j}\sin(tW_j^*)\sin(tW_k^*)\right\}\mathbb{I}(|t|\leq \kappa),
\end{equation}
where $\mathbb{I}(\cdot)$ is the indicator function and $\kappa$ is some positive constant. The constant $\kappa$ can be thought of as a smoothing parameter which ensures that the estimator $\hat{s_2}(t)$ behaves well for large values of $|t|$. Ideally, $\kappa$ can be chosen in a data-dependent way. Development of this approach is ongoing work. However, based on extensive simulation work, it has been found that values $\kappa \in [3,5]$ work reasonably well for a wide range of underlying GSS distributions considered. Now, taking equation (\ref{MISE}) and substituting $\hat{s_2}(t)$ for $s_0^2(t)$ and ignoring components that do not depend on the bandwidth gives approximate $\mathrm{MISE}$,
\begin{equation} \label{M_approx}
	\hat{M}(h) = \frac{1}{h} \int_{\mathbb R}\left\{ \frac{\psi_K^2(t)}{n\psi_U^2(t/(h\omega))} \left[\frac{1-\psi_U(2t/(h\omega))c_0(2t/h)}{2}\right] + \left(\frac{n-1}{n}\psi_K(t)-2\right)\psi_K(t)\hat{s_2}(t/h) \right\}dt.
\end{equation} 

The MISE-approximation bandwidth estimate is the value $\tilde{h}$ that minimizes $\hat{M}(h)$. The performance of both the CV and MISE-approximation bandwidth estimators will be investigated in Section \ref{sec:simulations} using a simulation study.

\section{Estimating the GSS Location and Scale Parameters} \label{sec:parm estim}

Up to this point, the location and scale parameters $\xi$ and $\omega$ have been treated as known. This is unrealistic in practice and therefore estimation of these parameters will be be considered. The problem of estimating the location and scale parameters of a GSS distribution with known symmetric component has received a great deal of attention in the literature for the non-measurement error setting, see \cite{ma2005locally}, \cite{azzalini2010invariance} and \cite{potgieter2013char}. However, this problem has not yet been considered in the presence of measurement error.

A Generalized Method of Moments (GMM) method for parameter estimation will be described in this section. Recall that $W_j = X_j + U_j = \xi + \omega Z_j + U_j$. Let $M\ge2$ be a positive integer and assume that both the GSS random variable $Z$ and the measurement error $U$ have at least $2M$ moments. Define
\begin{equation} \label{T_k def}
T_k := T_k\left(\xi,\omega\right) = n^{-1}\sum_{j=1}^{n}\left(\frac{W_j-\xi}{\omega}\right)^{2k}
\end{equation} 
with expectation
\begin{align} 
\mathrm{E}\left[T_k\right] & =\mathrm{E}\left[\left(Z+\omega^{-1}U\right)^{2k}\right]  \nonumber \\
& = \sum_{j=0}^{k} {{2k}\choose{2j}} \omega^{-2(k-j)} \mathrm{E} [Z^{2j}] \mathrm{E} [U^{2(k-j)}]. \label{E_Tk}
\end{align}
When evaluating equation \eqref{E_Tk}, let $Z_0$ have symmetric distribution with known pdf $f_0(z)$, corresponding the symmetric component of the GSS random variable $Z$. Then, by then property of even transformations (Proposition 1.4, Azzalini, 2013), $\mathrm{E} [Z^{2j}]=\mathrm{E} [Z_0^{2j}]$ for $j=1,\ldots,M$. Also, since the distribution of $U$ is known, the evaluation of its moments pose no problem. Thus, (numerical) evaluation of \eqref{E_Tk} is straightforward. Next, note that one can easily show that
\begin{equation} \label{E_TiTk}
\mathrm{E} \left[T_i T_k\right] = n^{-1}\mathrm{E} \left[T_{i+k} \right] + (n-1)n^{-1}\mathrm{E} \left[T_i\right]\mathrm{E} \left[T_k\right].
\end{equation}
Now, let
\begin{equation*}
\mathbf{T}_M=\left(T_1-\mathrm{E}[T_1],\ldots,T_M-\mathrm{E}[T_M]\right)
^\top\end{equation*}
and define covariance matrix $\mathbf{\Sigma}$ with entry in the $i$th row and $j$th column
\begin{equation*}
\mathbf{\Sigma}_{ij}=n^{-1}\left(\mathrm{E}\left[T_{i+j} \right]-\mathrm{E}\left[T_{i} \right]\mathrm{E}\left[T_{j} \right]\right).
\end{equation*}
Minimization of the quadratic form
\begin{equation} \label{min_dist}
D\left(\xi,\omega \right) = n \mathbf{T}_K^\top \mathbf{\Sigma}^{-1} \mathbf{T}_K
\end{equation}
gives the GMM estimators of the location and scale parameters. In evaluating $D(\xi,\omega)$, both the expectations $\mathrm{E}\left[T_{i} \right]$, $i=1,\ldots,M$ and the covariance matrix $\mathbf{\Sigma}$ are functions of the parameter $\omega$, but not of $\xi$. This method requires that both $Z$ and of $U$ have at least four finite moments, as the statistic $D$ is only defined for $M \ge 2$. In the case where $M=2$, minimization of $D$ is equivalent to method of moments with two equations in two unknowns.

There is one difficulty with the GSS estimator that needs to be pointed out. The statistic $D$ often has multiple minima. At first one might assume that the global minimum corresponds to the ``best'' solution. However, this equivalent problem also occurs in the non-measurement error setting when estimating the location and scale parameters of a GSS distribution with unknown skewing function. The solutions considered in the non-ME setting range from selecting the model with the least complex skewing function with complexity measure the squared integral of the second derivative of the function, to selecting the solution whose model-implied skewness is closest to the sample skewness, see Section 7.2.2 in \cite{Azzalini2013} for an overview and illustration. Additionally, simulation results for several GSS distributions and selection mechanisms can be found in \cite{potgieter2013char}. It is useful to further note that, as per \cite{Azzalini2013}, it is usually possible to select a most appropriate solution using a non-quantifiable approach such as visual inspection of the different estimated densities.  

For the problem at hand, assume that the quadratic form $D$ has $J$ local minima and let $(\hat{\xi}_j,\hat{\omega}_j)$, $j=1,\ldots,J$ denote the $J$ solution pairs obtained by minimizing $D$. Corresponding to the $j^{\mathrm{th}}$ solution, let $\tilde{f}_j(x|\hat{\xi}_j,\hat{\omega}_j)$ denote the GSS density deconvolution estimator with some suitably chosen bandwidth. Using this estimated density, let the $k$th implied moment associated with the $j$th solution be
\begin{equation} \label{implied_moments}
\tilde{\mu}_{j,k} = \int_{\mathbb{R}} x^k \tilde{f}_j(x|\hat{\xi}_j,\hat{\omega}_j) dx
\end{equation}
and the $j$th model-implied characteristic function
\begin{equation} \label{implied_cf}
\tilde{\phi}_j(t) = \int_{\mathbb{R}} \exp (itx) \tilde{f}_j(x|\hat{\xi}_j,\hat{\omega}_j) dx.
\end{equation}

In this paper, two different selection approaches are proposed, the first based on the underlying skewness of the distribution and the second based on the phase function of the distribution.

\textit{Method 1 (Skewness matching)}: For the model $W=X+U$, it is true that $\mathrm{Skew}(W)=(\sigma_X/\sigma_W)^3\mathrm{Skew}(X)$. Subsequently, an empirical estimate of the skewness of random variable $X$ is given by
\[\widehat{\mathrm{Skew}}(X)=\frac{\hat{\sigma}_W^2}{(\hat{\sigma}_W^2-{\sigma}_U^2)^{3/2}}\widehat{\mathrm{Skew}}(W)\]
where $\hat{\sigma}_W^2$ and $\widehat{\mathrm{Skew}}(W)$ denote the sample variance and skewness of $W$. Now, for the $j$th solution pair $(\hat{\xi}_j,\hat{\omega}_j)$, the model-implied skewness is given by \[\hat{\gamma}_j=\frac{\tilde{\mu}_{j,3}-3\tilde{\mu}_{j,2}\tilde{\mu}_{j,1}+2\tilde{\mu}_{j,1}^3}{\tilde{\mu}_{j,2}-\tilde{\mu}_{j,1}^2}\] with $\tilde{\mu}_{j,k}$ defined in \eqref{implied_moments}. The selected solution is the one with implied skewness closest to the empirical skewness. Specifically, let $d_j = |\widehat{\mathrm{Skew}}(X)-\hat{\gamma}_j|$, $j=1,\ldots,J$, then the selected solution corresponds to index $j^* = \arg\min_{1\leq j \leq J} d_j$.

\textit{Method 2 (Phase function distance)}: The empirical phase function was recently used by \cite{delaigle2016methodology} for density deconvolution where the measurement error is symmetric but of unknown type. The phase function, defined as the ratio of the characteristic function and its norm, is invariant to the addition of measurement error provided the distribution of the measurement error is symmetric about $0$. Specifically, for the present model $W=X+U$, let $\rho_W(t)$ and $\rho_X(t)$ denotes the phase functions associated with random variables $W$ and $X$, then $\rho_W(t)=\rho_X(t)$ for all $t$. Now, the empirical estimate of the phase function of $X$ is $\hat{\rho}_X(t)=\hat{\psi}_W(t)/|\hat{\psi}_W(t)|$ where $|z|=(z\bar{z})^{1/2}$ is norm of complex number $z$ with $\bar{z}$ denoting the complex conjugate of $z$. For the $j$th solution, the model-implied phase function is given by $\tilde{\rho}_j(t)=\tilde{\phi}_j(t)/|\tilde{\phi}_j(t)|$ with $\tilde{\phi}_j(t)$ defined in \eqref{implied_cf}. Let $w(t)$ denote a non-negative weight function symmetric around zero and define $j$th phase function distance
\[R_j = \int_{\mathbb{R}} \vert \hat{\rho}_X(t) - \tilde{\rho}_j(t) \vert w(t) dt.\] The selection solution is the one with the smallest phase function distance $R_j$.

The deconvolution and selection procedure can thus be approaches as follows. For the $j$th solution pair $(\hat{\xi}_j,\hat{\omega}_j)$:
\begin{itemize}
	\item calculate values $\hat{W}^{\ast}_{ij}=(W_i-\hat{\xi}_j)/\hat{\omega}_j$ for $i=1,\ldots,n$;
	\item estimate a bandwidth $\tilde{h}_j$ using data $\hat{W}^{\ast}_{ij}$;
	\item estimate skewing function $\tilde{\pi}_j(z)$ using equations (\ref{pi_est1}) and (\ref{pi_est2});
	\item calculate deconvolution density estimate $\tilde{f}_j(x|\hat{\xi}_j,\hat{\omega}_j)$ using equation (\ref{f_tilde_X});
	\item calculate selection criteria according to either the skewness matching or phase function distance methods.
\end{itemize}
Finally, select estimators $(\hat{\xi},\hat{\omega})=(\hat{\xi}_{j^*},\hat{\omega}_{j^*})$ where index $j^{\ast}$ corresponds to the selected solution according to the criterion used and define the estimated GSS deconvolution density to be $\tilde{f}_X(x)=\tilde{f}_{j^*}(x|\hat{\xi}_{j^*},\hat{\omega}_{j^*})$.

\section{Simulation Studies} \label{sec:simulations}

\begin{figure}[!h]
	\begin{center}
		\includegraphics[scale=0.8]{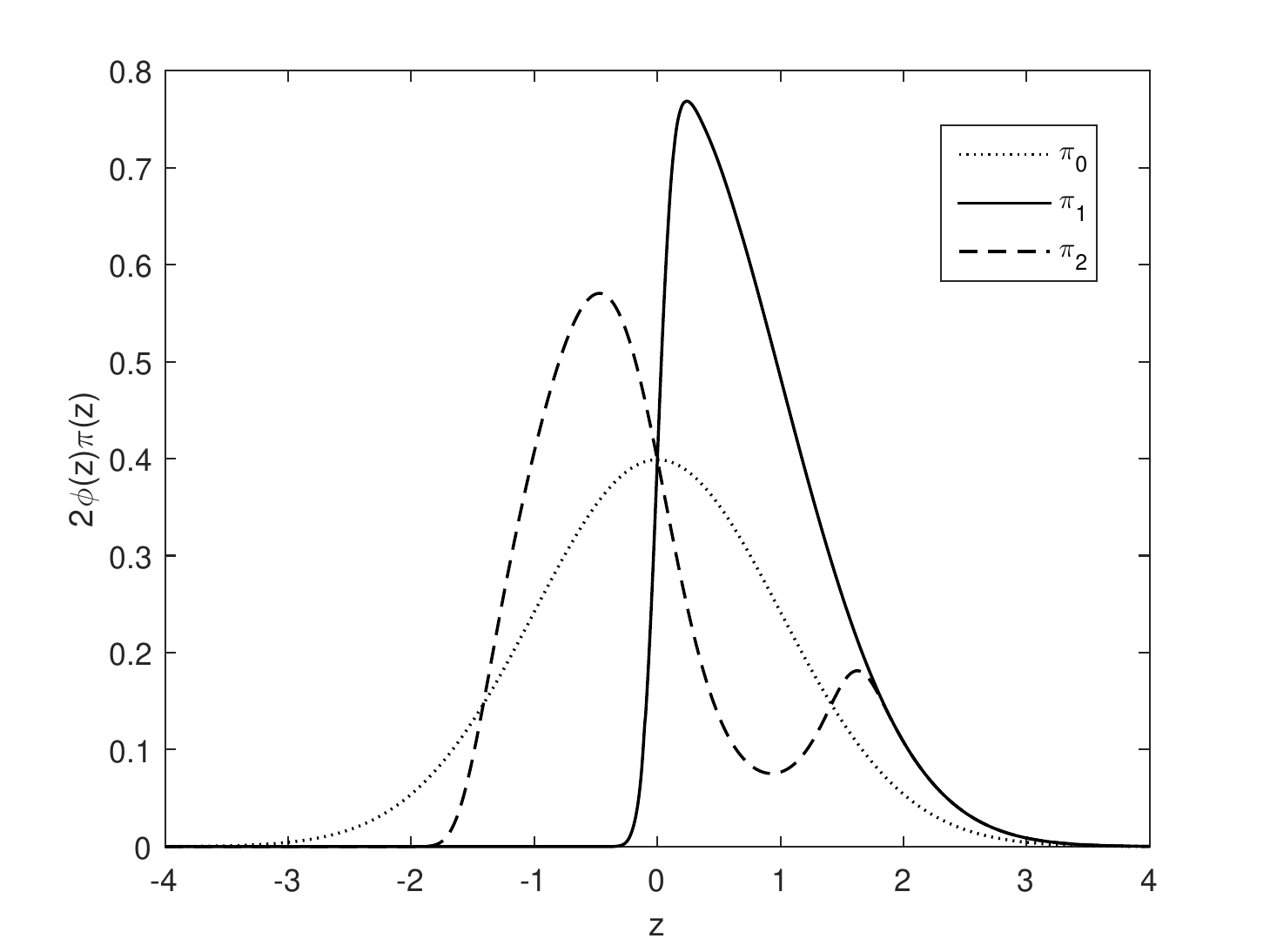}
		\caption{Skew-symmetric densities used in simulation study.}
		\label{fig:three_dens}
	\end{center}
\end{figure}

Several simulation studies were done to investigate the performance of the GSS deconvolution estimator. The simulations investigated the GMM parameter estimation method, the performance of the GSS and nonparametric deconvolution estimators under optimal conditions, the proposed CV and MISE-approximation bandwidth selection methods, and the outlined solution selection algorithm. In all these simulation studies, samples $Z_1,\ldots,Z_n$ were generated from a GSS distribution with normal symmetric component, $f_0(z)=\phi(z)$ and with three different skewing functions, namely $\pi_0(z)=0.5$, $\pi_1(z)=\Phi(9.9625z)$ and $\pi_2(z)=\Phi(z^3-2z)$ where $\phi$ and $\Phi$ are the standard normal density and distribution functions. The location and scale parameters were taken to be $\xi=0$ and $\omega=1$, so that $X_j=Z_j$ for all $j$, while two measurement error scenarios were considered, namely $U_j$ following either a normal or a Laplace distributions with variances chosen so that the noise-to-signal ratio $\mathrm{NSR}=\sigma_U^2/\sigma_X^2 \in \{0.2,0.5\}$. Samples of size $n=200$ and $500$ of observations $W_j = X_j + U_j$, $j=1,\ldots,n$ were generated from each configuration of skewing function, measurement error distribution and $\mathrm{NSR}$.Figure \ref{fig:three_dens} illustrates the diversity of shapes of the density functions $2\phi(z)\pi_j(z)$, $j=0,1,2$. The skewing function $\pi_0$ recovers the normal distribution. The skewing function $\pi_1$ results in a positive skew distribution, while $\pi_2$ results in a bimodal distribution.

\begin{center}
	Figure \ref{fig:three_dens} About Here
\end{center}

The first simulation done considered estimation of $(\xi,\omega)$ using the GMM method. The simulation compared the estimators obtained by minimizing (\ref{min_dist}) for $M=2$ and $M=5$ even moments. The goal of this simulation was to determine whether ``more'' information (the use of additional sample moments) results in better estimators using RMSE as a criterion. In this simulation, estimators using only the second and fourth moments are compared to estimators using even moments up to and including the tenth moment. While the sixth, eight and tenth moments arguably contain additional information, there is a great deal of added variability introduced when estimating these from the sample. This simulation considers a ``best case'' scenario in that when there are multiple solutions $(\hat{\xi}_j,\hat{\omega}_j)$, the solution closest to the true value $(0,1)$ as measured using Euclidean distance is selected. A total of $N=1000$ samples were drawn from each simulation configuration. The results are shown in Table \ref{tab:GMM_estims} below.

\begin{center}
	Table \ref{tab:GMM_estims} About Here
\end{center}

Several observations can be made upon inspection of Table \ref{tab:GMM_estims}. Consider the simulations in the setting $\pi(z)=\pi_0(z)$, i.e. the distribution of $X$ is normal. In most instances, the use of $M=5$ moments results in a small increase in RMSE compared to the case $M=2$ when considering the estimates. The average increase in RMSE for $\xi$ is $1.2\%$ and for $\omega$ is $9.5\%$ across the settings considered. On the other hand, the simulation results for skewing functions $\pi_1(z)$ and $\pi_2(z)$ look very different. For each simulation configuration, there was a large decrease in RMSE for $\xi$ and a large decrease in the RMSE of $\omega$ for skewing function $\pi_1(z)$. For the skewing function $\pi_1(z)$, the average decrease in RMSE is $27.7\%$ for $\xi$ and $18.5\%$ for $\omega$. For the skewing function $\pi_2(z)$, the average decrease in RMSE is $32\%$ for $\xi$, but on average the RMSE for $\omega$ remains unchanged across the simulation settings considered. One possible reason for the increase in RMSE in the $\pi_0$-case is that the underlying distribution is normal and therefore higher-order moments do not contain ``extra'' information about the distribution. On the other hand, the $\pi_1$ and $\pi_2$ cases depart substantially from normality and the higher-order sample moments, despite their large variability, do contain information about the underlying distribution. As the increase in RMSE in the symmetric case is relatively small compared to the decrease in the asymmetric cases, for the remainder of this paper the GMM estimators with $M=5$ will be used to estimate $(\xi,\omega)$.

Simulation studies were also done to compare the proposed GSS deconvolution estimator to the established nonparametric deconvolution estimator. Specifically, a simulation study was done to compare the performance of the two estimators assuming that in each instance the bandwidth could be chosen to minimize the true ISE. For a sample $W_1,\ldots,W_n$, let $\tilde{f}_{\mathrm{GSS}}(x|h)$ and $\tilde{f}_{\mathrm{NP}}(x|h)$ denote, respectively, the GSS estimator and the nonparametric estimator; in both instances $h$ denotes the bandwidth parameter. The ISE is defined as
\[\mathrm{ISE}_{est}(h) = \int_{\mathbb R} \left[\tilde{f}_{est}(x|h)-f_X(x)\right]^2dx\]
where $est \in \{\mathrm{GSS},\mathrm{NP}\}$. For each simulated sample, the bandwidth minimizing ISE was found for both the GSS and nonparametric estimators. For the GSS estimator, when the GMM gave multiple possible solutions for $(\xi,\omega)$, the solution with smallest ISE was chosen. The results summarized in Table \ref{tab:ISE_opt_estims} therefore represent the performance of the two estimators if one could choose the bandwidth minimizing ISE and then choose the GSS solution with smallest ISE. While this is not doable in practice, it is useful to compare the estimators under such idealized conditions, as it speaks to their best possible performance. For each simulation configuration, $N=1000$ samples were generated. Due to the occasional occurrence of very large outliers in ISE, the median ISE rather than the mean ISE is reported. Additionally, the first and third quartiles of ISE are also reported.

\begin{center}
	Table \ref{tab:ISE_opt_estims} About Here
\end{center}

Under the optimal bandwidth selection scenario considered, inspection of Table \ref{tab:ISE_opt_estims} shows how well the GSS deconvolution estimator can perform relative to the nonparametric deconvolution estimator. In the symmetric case ($\pi_0$), the reduction in median ISE exceeds $50\%$ in all cases. The reduction in median ISE is most dramatic in the symmetric case. However, for the other two skewing functions ($\pi_1, \pi_2$), the reduction in median ISE is seen to be as large as $40\%$. There is one instance where median ISE of the nonparametric estimator is smaller than that of the GSS estimator -- skewing function $\pi_2$, $NSR=0.5$, Laplace measurement error and sample size $n=200$. However, the equivalent scenario with sample size $n=500$ has the GSS estimator with smaller median ISE again. This likely indicates the effect of the variability of estimating the location and scale parameters in a small sample, especially when large amounts of heavier-tailed-than-normal measurement error is present. Overall, the good performance indicated in these results help motivate the study of the GSS deconvolution estimator. While there is some additional structure being imposed by the GSS estimator (the a priori specification of the symmetric density $f_0$), there are potentially large decreases in median ISE.

Next, an extensive simulation study was done looking at the two proposed bandwidth estimation methods together with the solution selection methods. For each simulated sample, the CV and MISE-approximation bandwidths were selected for each possible GMM solution $(\hat{\xi}_j,\hat{\omega}_j)$. Additionally, the two-stage plug-in bandwidth of \cite{delaigle2002estimation}, developed for nonparametric deconvolution, was also included to compare its performance in when applied in the GSS setting. Bandwidths are always estimated based on the transformed data $\hat{W}^\ast_{ij}=(W_{ij}-\hat{\xi}_j)/\hat{\omega}_j$. After a bandwidth was selected for each solution, both the skewness matching and phase function distance metric selection methods were implemented in order to choose between multiple solutions. To contextualize these results, the solution with smallest ISE was chosen to represent the ``best possible'' performance, while blind selection was also implemented by randomly selecting one of the solutions. Finally, the nonparametric deconvolution estimator with two-stage plug-in bandwidth was calculated for reference purposes. Most of these simulation results are summarized in Tables A1 through A5 in the Supplemental Material, but one of these tables is included here for illustration. Table \ref{tab:ISE_MISE500_selection} reports the median, as well as first and third quartiles, of ISE for $N=1000$ simulated datasets with sample size $n=500$, MISE-approximation bandwidth estimation and all selection criteria mentioned.

\begin{center}
	Table \ref{tab:ISE_MISE500_selection} About Here
\end{center}

Inspection of Table \ref{tab:ISE_MISE500_selection} shows that, in the case of the MISE-approximation bandwidth, both the skewness and phase function selection approaches generally perform better than the usual nonparametric estimator, the exception being the combination of skewing function $\pi_2$ and Laplace measurement error. As the GSS estimator outperformed the nonparametric estimator under ``optimal'' bandwidth selection in Table \ref{tab:ISE_opt_estims}, this does suggest that one might still be able to improve performance of the GSS estimator by some combination of improved parameter estimation and bandwidth selection -- this is ongoing work. Further inspection of Table \ref{tab:ISE_MISE500_selection} shows that both the skewness and phase function selection mechanisms generally perform better than random selection, with the exception that random selection outperforms the skewness approach for skewing function $\pi_1$ and normal measurement error. While there are a few instances where skewness-based selection outperforms phase function-based selection, the latter generally has very good performance and comes close to the best possible performance of the minimum ISE. 

Inspection of Tables A1 through A5 lead to a general conclusion: regardless of the bandwidth estimation method, phase function-based selection tends to performs better than skewness-based selection. As such, the median ISE values for the three bandwidth estimation methods considered used together with phase function-based selection are summarized below in Tables \ref{tab:medianISE_estims200} and \ref{tab:medianISE_estims500}. The nonparametric estimator performance is again included for reference purposes.

\begin{center}
	Tables \ref{tab:medianISE_estims200} and \ref{tab:medianISE_estims500} About Here
\end{center}

In Tables \ref{tab:medianISE_estims200} and \ref{tab:medianISE_estims500}, the CV bandwidth method performs poorly, having larger median ISE than the MISE-approximation and two-stage plug-in methods for skewing function $\pi_1$ and $\pi_2$. However, in the symmetric case ($\pi_0$), the CV method does tend to outperform the MISE method. For the underlying symmetric distribution ($\pi_0$) and bimodal distribution ($\pi_2$), the two-stage plug-in method has the best performance, beating the nonparametric estimator except for the underlying bimodal distribution with Laplace measurement error and $NSR=0.5$. For the underlying unimodal skew distribution ($\pi_1$), the MISE-approximation bandwidth has best performance, beating the nonparametric estimator in this case. In most simulation settings, both the MISE-approximation and two-stage plug-in bandwidth methods combined with phase function-based selection result in better performance than the nonparametric estimator, except for the bimodal distribution with large measurement error variance, i.e. when $NSR=0.5$.

\section{Application}
\subsection{Coal Abrasiveness Index Data}
\begin{figure}[!h]
	\begin{center}
		\includegraphics[scale=0.6]{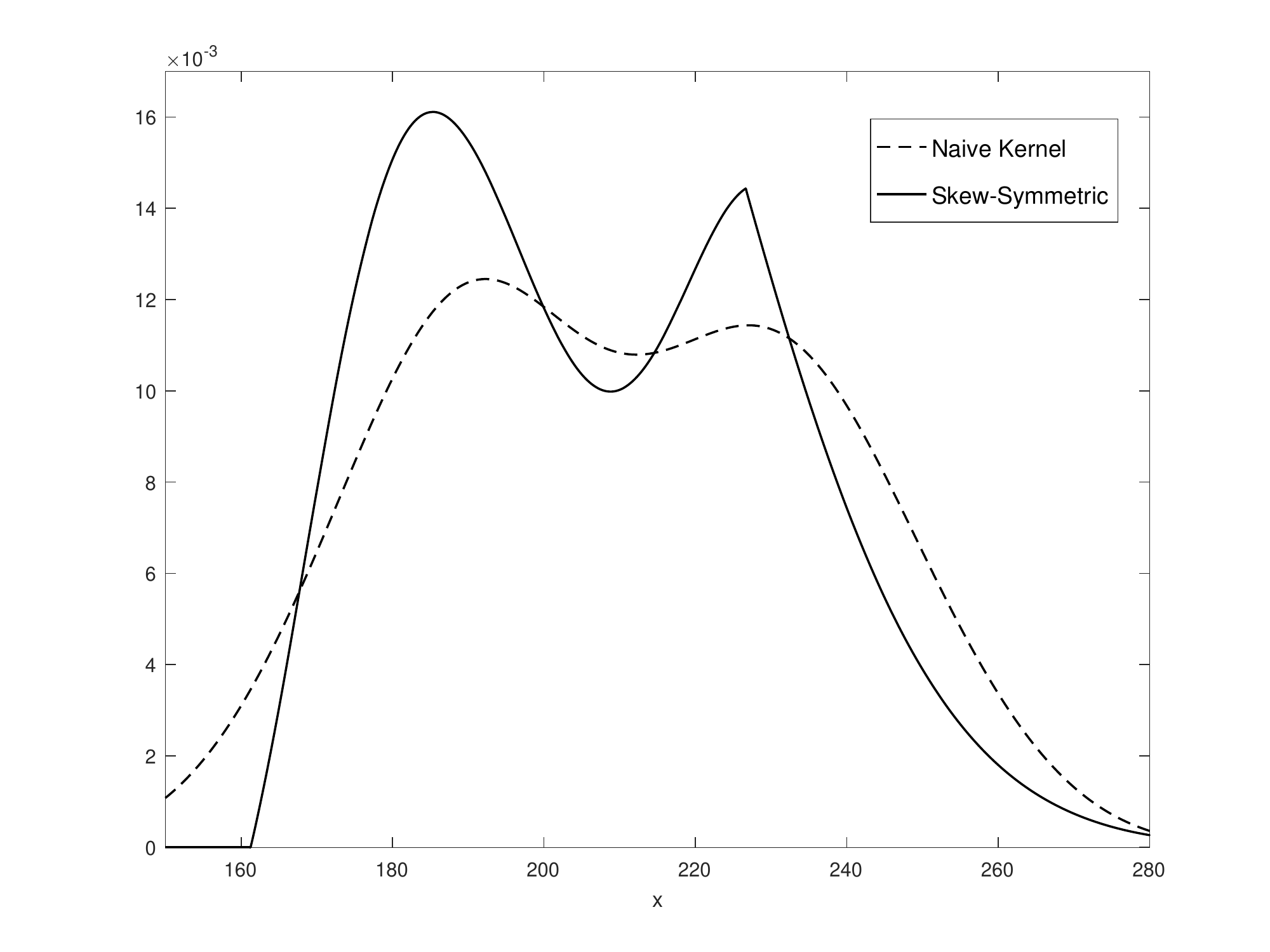}
		\caption{Abrasiveness Index Density Estimation.}
		\label{fig:AI_decon}
	\end{center}
\end{figure}

The data analyzed here are from an industrial application and were first considered by \cite{lombard2005nonparametric}. The data were obtained by taking batches of coal, splitting them in two, and randomly allocating each of the two half-batches to one of two methods used to measure the abrasiveness index (AI) of coal. The AI is a measure of the quality of the coal. The data consist of $98$ pairs $\left( W_{1i},W_{2i}\right) $ where it is assumed that $W_{1i}=X_{i}+U_{1i}$
and $W_{2i}=\mu +\sigma \left( X_{i}+U_{2i}\right) $ where $X_i$ denotes the true AI of the $i$th batch, $U_{1i}$ and $U_{i2}$ denote measurement error, and the constants $\mu$ and $\sigma$ are location and scale parameters used to account for the two methods measuring the AI on different scales. These variables have first and second moments $\mu_{W_1} = \mu_X$, $\mu_{W_2} = \sigma\mu_X$,  $\sigma _{W_{1}}^{2}=\sigma _{X}^{2}+\sigma _{U}^{2}$, and $\sigma _{W_{2}}^{2}=\sigma ^{2}\left( \sigma _{X}^{2}+\sigma _{U}^{2}\right)$. By replacing the population moments with their sample equivalents and solving, estimators $\hat{\sigma}=S_{W_2}/S_{W_1}=0.679$ and $\hat{\mu}=\bar{W}_2 - \hat{\sigma}\bar{W}_1=59.503$ are observed with $(\bar{W}_1,S_{W_1})$ denoting the sample mean and standard deviation of the $W_1$-data and similar definitions holding for the $W_2$-quantities. Now, define
\[
	W_{i} =\frac{1}{2}W_{1i}+\frac{1}{2}\left( \frac{W_{2i}-\hat{\mu}}{\hat{%
			\sigma}}\right) \]
and note that \[
	W_i \approx X_i+\frac{1}{2}\left( U_{1i}+U_{2i}\right)
	=X_i+\varepsilon_i.
\]
An estimate of the measurement error variance $\sigma_\varepsilon^2$ can be obtained by calculating
\[
\hat{\sigma}_{U}^{2}=\frac{1}{2n}\sum \left( W_{1i}-\frac{W_{2i}-\hat{\mu}}{	\hat{\sigma}} \right) ^{2}=174.6
\]
and noting that
\[
\hat{\sigma}_{\varepsilon }^{2}=\frac{174.6}{2}=87.3
\]%
which corresponds to the $W_i$ having noise-to-signal ratio $NSR=16.35\%$. The GSS deconvolution estimator for the true AI, $f_X(x)$, is calculated assuming a normal symmetric component $f_0(z)$, and a Laplace distribution for the measurement error $\varepsilon$. Using the GMM approach with $M=5$, two possible solutions pairs are found, namely $(\hat{\xi}%
_{1},\hat{\omega}_{1}) =\left( 192.88,29.90\right) $ and $(\hat{\xi}_{2},\hat{\omega}_{2}) =\left( 230.41,32.43\right) $. For
each solution a corresponding skewing function $\tilde{\pi}_{j}\left( z\right)$ was
estimated and then the phase function distance statistic $R_j$ was calculated using weight function $w(t)=(1-(t/t^\ast)^2)^3$ for $t\in [-t^\ast,t^\ast]$ and $t^\ast=0.06$ in this application. This gave phase function distance statistics $%
R_{1}=0.023<0.046=R_{2}$ and subsequently the solution $(\hat{\xi}%
_{1},\hat{\omega}_{1},\tilde{\pi}_1)$ was selected. Skewness matching resulted in selection of the same solution. The MISE-approximation bandwidth for this method was $\tilde{h}%
=0.102$. Figure \ref{fig:AI_decon} shows a kernel density estimator of the
contaminated measurements $W$ as well as the GSS deconvolution estimator of $f_X$. 

\begin{center}
	Figure \ref{fig:AI_decon} About Here
\end{center}

This application illustrates one of the downsides of the GSS approach in smaller samples. Note the sharp edge in the GSS deconvolution density estimator around $x=225$. This is an artifact of the hard truncation applied when estimating the skewing function in \eqref{pi_est2}. The estimator is not differentiable at points such as this one and are, as such, equivalent to points where the nonparametric kernel estimator is not differentiable because it has been truncated to be positive.

The GSS deconvolution estimator was also calculated assuming normally distributed measurement error, the results were nearly identical.

\subsection{Systolic Blood Pressure Application}

The data here are a subset of $n=1615$ observations from the Framingham Heart Study. All study participants in the subset are men. The dataset includes systolic blood pressure measured twice at both the second and third patient exams, $SBP_{21},SBP_{22},SBP_{31}$ and $SBP_{32}$. Measurement of systolic blood pressure is subject to a large amount of measurement error. As suggested by \cite{carroll2006measurement}, define $P_1 = (SBP_{21}+SBP_{22})/2 $ and $P_2 = (SBP_{31}+SBP_{32})/2 $ to be the average systolic blood pressure observed at each of exams two and three. The transformed variables $W_j = \log (P_j - 50)$, $j=1,2$ are then calculated to adjust for large skewness present in the data. The measurement $W=(W_1+W_2)/2$ is a surrogate for the true long-term average systolic blood pressure (on the transformed logarithmic scale) $X$. Using the replicate measurements $W_1$ and $W_2$, we are able to estimate standard deviations $\hat{\sigma}_X=0.1976$ and $\hat{\sigma}_U=0.0802$ in the relationship $W=X+U$.

\begin{figure}[h!]
	\begin{center}
	\includegraphics[width=1.1\textwidth]{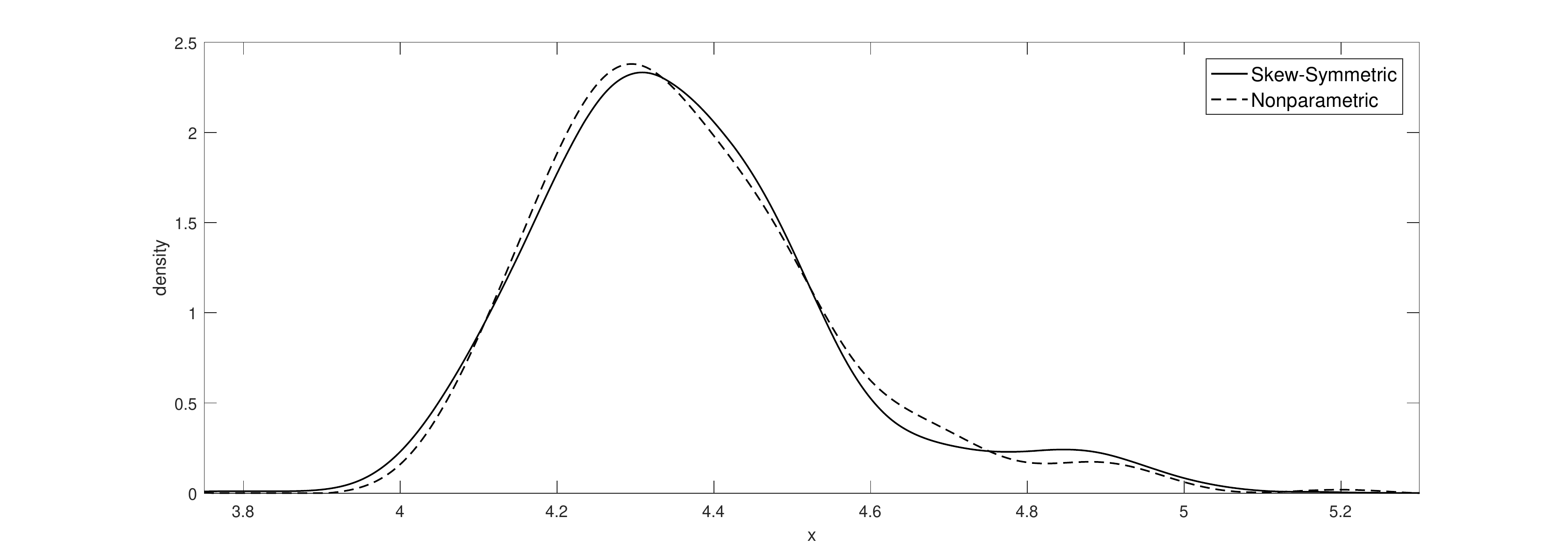}
	\caption{Density deconvolution estimators of log(SBP-50)}
	\label{fig:SBP_decon}
	\end{center}
\end{figure}

Assuming that the measurement error follows a Laplace distribution and that the symmetric density $f_0(z)$ is normal, implementation of the GMM method with $M=5$ gives location and scale estimates $(\hat{\xi},\hat{\omega})=(4.429,0.210)$. As there is only one solution, no selection criterion needs to be used. Using the transformed data $\hat{W}^{*}=(W-4.429)/0.21$, the two-stage plug-in bandwidth of \cite{delaigle2002estimation} was calculated, $\tilde{h} = 0.119$. Figure \ref{fig:SBP_decon} displays both the GSS density deconvolution estimator of (\ref{f_tilde_X}) as well as the frequently used nonparametric kernel deconvolution density estimator, also with two-stage plug-in bandwidth.

\begin{center}
	Figure \ref{fig:SBP_decon} About Here
\end{center}

The nonparametric kernel estimator has previously been used in the Framingham Heart Study. In this particular application, it is reassuring that the GSS estimator is not dissimilar from the nonparametric estimator.

\section{Conclusion}

This paper develops a density deconvolution approach assuming the density of interest is a member of the generalized skew-symmetric (GSS) family of distributions with known symmetric component. In practice, calculation of this deconvolution estimator requires both the estimation of location and scale parameters $(\xi,\omega)$, as well as the estimation of a skewing function $\pi(z)$. The skewing function estimator is nonparametric in nature and typically has a slow rate of convergence depending on the distribution of the measurement error, but the location and scale parameter estimators are obtained using a method of moments approach and converge at the usual root-$n$ parametric rate. The effect of estimating these parameters therefore becomes negligible in large samples relative to the variability in estimating the skewing function. 

The skewing function estimator depends on a bandwidth parameter. Two approaches are developed for bandwidth estimation, one a cross-validation type method and the other an approximation to the MISE. These methods are compared against the two-stage plug-in bandwidth of \cite{delaigle2002estimation} which was developed for nonparametric deconvolution. Based on simulation studies carried out, the MISE-approximation bandwidth and two-stage plug-in bandwidths are seen to perform better than cross-validation bandwidth. The good performance of the two-stage plug-in approach opens up one avenue for future research; the development of a plug-in bandwidth selection approach specific to the GSS framework is currently being investigated.

One complication sometimes encountered in the GSS setting is the need to choose between competing estimators for $(\xi,\omega)$. This equivalent problem also occurs in the non-measurement error setting for GSS distributions. Two methods are proposed for doing this selection, one based on the sample skewness, and a second based on the empirical phase function. Extensive simulations are done and the phase function approach is seen to usually have better performance than the skewness approach.

While a combination of either the MISE-approximation or two-stage plug-in bandwidths together with phase function selection perform very well in simulations, there were a few instances where the nonparametric estimator had superior performance. This suggests that improvements to either the parameter estimates and/or the bandwidth selection might be possible. This is also a current avenue of research being pursued.

\bibliographystyle{biomAbhra}
\bibliography{Potgieter_2017}

\newpage

\section*{Figures and Tables}

\begin{table}[!h]
\begin{center}
\begin{tabular}{ccccccc}
	\cline{4-7}
	&  &  & \multicolumn{2}{c}{$M=2$} & \multicolumn{2}{c}{$M=5$} \\ \hline
	\multicolumn{1}{c}{$\pi $} & \multicolumn{1}{c}{$n$} & 
	\multicolumn{1}{c}{$\left( \mathrm{NSR},U\right) $} & $\mathrm{RMSE}(\hat{%
		\xi})$ & $\mathrm{RMSE}(\hat{\omega})$ & $\mathrm{RMSE}(\hat{\xi})$ & $%
	\mathrm{RMSE}(\hat{\omega})$ \\ \hline
	\multicolumn{1}{c}{$\pi _{0}$} & \multicolumn{1}{c}{$200$} & 
	\multicolumn{1}{c}{$\left( 0.2,N\right) $} & $0.400$ & $0.116$ & $0.404$ & 
	$0.127$ \\ 
	\multicolumn{1}{c}{} & \multicolumn{1}{c}{} & \multicolumn{1}{c}{$\left(
		0.5,N\right) $} & $0.454$ & $0.140$ & $0.452$ & $0.153$ \\ 
	\multicolumn{1}{c}{} & \multicolumn{1}{c}{} & \multicolumn{1}{c}{$\left(
		0.2,L\right) $} & $0.409$ & $0.120$ & $0.414$ & $0.133$ \\ 
	\multicolumn{1}{c}{} & \multicolumn{1}{c}{} & \multicolumn{1}{c}{$\left(
		0.5,L\right) $} & $0.494$ & $0.157$ & $0.483$ & $0.168$ \\ \cline{2-7}
	\multicolumn{1}{c}{} & \multicolumn{1}{c}{$500$} & \multicolumn{1}{c}{$%
		\left( 0.2,N\right) $} & $0.370$ & $0.094$ & $0.383$ & $0.105$ \\ 
	\multicolumn{1}{c}{} & \multicolumn{1}{c}{} & \multicolumn{1}{c}{$\left(
		0.5,N\right) $} & $0.415$ & $0.113$ & $0.431$ & $0.128$ \\ 
	\multicolumn{1}{c}{} & \multicolumn{1}{c}{} & \multicolumn{1}{c}{$\left(
		0.2,L\right) $} & $0.377$ & $0.097$ & $0.389$ & $0.109$ \\ 
	\multicolumn{1}{c}{} & \multicolumn{1}{c}{} & \multicolumn{1}{c}{$\left(
		0.5,L\right) $} & $0.453$ & $0.133$ & $0.453$ & $0.136$ \\ \hline
	\multicolumn{1}{c}{$\pi _{1}$} & \multicolumn{1}{c}{$200$} & 
	\multicolumn{1}{c}{$\left( 0.2,N\right) $} & $0.131$ & $0.112$ & $0.092$ & 
	$0.091$ \\ 
	\multicolumn{1}{c}{} & \multicolumn{1}{c}{} & \multicolumn{1}{c}{$\left(
		0.5,N\right) $} & $0.177$ & $0.138$ & $0.151$ & $0.121$ \\ 
	\multicolumn{1}{c}{} & \multicolumn{1}{c}{} & \multicolumn{1}{c}{$\left(
		0.2,L\right) $} & $0.139$ & $0.117$ & $0.092$ & $0.093$ \\ 
	\multicolumn{1}{c}{} & \multicolumn{1}{c}{} & \multicolumn{1}{c}{$\left(
		0.5,L\right) $} & $0.195$ & $0.154$ & $0.152$ & $0.124$ \\ \cline{2-7}
	\multicolumn{1}{c}{} & \multicolumn{1}{c}{$500$} & \multicolumn{1}{c}{$%
		\left( 0.2,N\right) $} & $0.080$ & $0.069$ & $0.055$ & $0.057$ \\ 
	\multicolumn{1}{c}{} & \multicolumn{1}{c}{} & \multicolumn{1}{c}{$\left(
		0.5,N\right) $} & $0.103$ & $0.084$ & $0.079$ & $0.071$ \\ 
	\multicolumn{1}{c}{} & \multicolumn{1}{c}{} & \multicolumn{1}{c}{$\left(
		0.2,L\right) $} & $0.083$ & $0.072$ & $0.055$ & $0.058$ \\ 
	\multicolumn{1}{c}{} & \multicolumn{1}{c}{} & \multicolumn{1}{c}{$\left(
		0.5,L\right) $} & $0.118$ & $0.097$ & $0.079$ & $0.073$ \\ \hline
	\multicolumn{1}{c}{$\pi _{2}$} & \multicolumn{1}{c}{$200$} & 
	\multicolumn{1}{c}{$\left( 0.2,N\right) $} & $0.133$ & $0.055$ & $0.096$ & 
	$0.058$ \\ 
	\multicolumn{1}{c}{} & \multicolumn{1}{c}{} & \multicolumn{1}{c}{$\left(
		0.5,N\right) $} & $0.234$ & $0.071$ & $0.185$ & $0.068$ \\
	\multicolumn{1}{c}{} & \multicolumn{1}{c}{} & \multicolumn{1}{c}{$\left(
		0.2,L\right) $} & $0.153$ & $0.058$ & $0.093$ & $0.059$ \\ 
	\multicolumn{1}{c}{} & \multicolumn{1}{c}{} & \multicolumn{1}{c}{$\left(
		0.5,L\right) $} & $0.334$ & $0.109$ & $0.194$ & $0.088$ \\ \cline{2-7}
	\multicolumn{1}{c}{} & \multicolumn{1}{c}{$500$} & \multicolumn{1}{c}{$%
		\left( 0.2,N\right) $} & $0.081$ & $0.034$ & $0.059$ & $0.037$ \\ 
	\multicolumn{1}{c}{} & \multicolumn{1}{c}{} & \multicolumn{1}{c}{$\left(
		0.5,N\right) $} & $0.135$ & $0.037$ & $0.112$ & $0.039$ \\ 
	\multicolumn{1}{c}{} & \multicolumn{1}{c}{} & \multicolumn{1}{c}{$\left(
		0.2,L\right) $} & $0.093$ & $0.035$ & $0.057$ & $0.037$ \\ 
	\multicolumn{1}{c}{} & \multicolumn{1}{c}{} & \multicolumn{1}{c}{$\left(
		0.5,L\right) $} & $0.219$ & $0.061$ & $0.124$ & $0.054$ \\ \hline
\end{tabular}
\caption{RMSE for GMM estimation $(\xi,\omega)$ with $M = 2,5$ and different simulation configurations, $N=\mathrm{Normal}$, $L=\mathrm{Laplace}$.}
\label{tab:GMM_estims}
\end{center}
\end{table}

\begin{table}[!h]
	\begin{center}
\begin{tabular}{ccccccc}
	\cline{3-7}
	&  & \multicolumn{2}{c}{$n=200$} &  & \multicolumn{2}{c}{$n=500$} \\ \hline
	$\pi $ & $\left( \mathrm{NSR},U\right) $ & $\mathrm{GSS}$ & $\mathrm{NP}$ & 
	& $\mathrm{GSS}$ & $\mathrm{NP}$ \\ \hline
	$\pi _{0}$ & $\left( 0.2,N\right) $ & $0.131$ & $0.442$ &  & $0.070$ & $0.282
	$ \\ 
	&  & $\left[ 0.055,0.263\right] $ & $\left[ 0.256,0.709\right] $ &  & $\left[
	0.032,0.148\right] $ & $\left[ 0.186,0.418\right] $ \\ 
	& $\left( 0.5,N\right) $ & $0.199$ & $0.817$ &  & $0.122$ & $0.596$ \\ 
	&  & $\left[ 0.084,0.405\right] $ & $\left[ 0.532,1.228\right] $ &  & $\left[
	0.048,0.296\right] $ & $\left[ 0.409,0.845\right] $ \\ 
	& $\left( 0.2,L\right) $ & $0.113$ & $0.273$ &  & $0.058$ & $0.147$ \\ 
	&  & $\left[ 0.053,0.241\right] $ & $\left[ 0.140,0.476\right] $ &  & $\left[
	0.027,0.117\right] $ & $\left[ 0.079,0.236\right] $ \\ 
	& $\left( 0.5,L\right) $ & $0.148$ & $0.327$ &  & $0.076$ & $0.169$ \\ 
	&  & $\left[ 0.074,0.323\right] $ & $\left[ 0.165,0.603\right] $ &  & $\left[
	0.040,0.158\right] $ & $\left[ 0.086,0.308\right] $ \\ \hline
	$\pi _{1}$ & $\left( 0.2,N\right) $ & $1.690$ & $2.453$ &  & $1.400$ & $1.875
	$ \\ 
	&  & $\left[ 1.271,2.188\right] $ & $\left[ 1.855,3.173\right] $ &  & $\left[
	1.031,1.775\right] $ & $\left[ 1.434,2.419\right] $ \\ 
	& $\left( 0.5,N\right) $ & $2.277$ & $4.079$ &  & $2.034$ & $3.514$ \\ 
	&  & $\left[ 1.729,2.956\right] $ & $\left[ 3.116,5.275\right] $ &  & $\left[
	1.547,2.645\right] $ & $\left[ 2.716,4.352\right] $ \\ 
	& $\left( 0.2,L\right) $ & $1.200$ & $1.701$ &  & $0.712$ & $1.096$ \\ 
	&  & $\left[ 0.832,1.658\right] $ & $\left[ 1.223,2.258\right] $ &  & $\left[
	0.422,1.112\right] $ & $\left[ 0.818,1.463\right] $ \\ 
	& $\left( 0.5,L\right) $ & $1.542$ & $2.353$ &  & $1.025$ & $1.615$ \\ 
	&  & $\left[ 1.054,2.162\right] $ & $\left[ 1.671,3.176\right] $ &  & $\left[
	0.652,1.469\right] $ & $\left[ 1.206,2.105\right] $ \\ \hline
	$\pi _{2}$ & $\left( 0.2,N\right) $ & $1.410$ & $1.768$ &  & $1.004$ & $1.289
	$ \\ 
	&  & $\left[ 0.918,2.082\right] $ & $\left[ 1.251,2.465\right] $ &  & $\left[
	0.689,1.406\right] $ & $\left[ 0.971,1.719\right] $ \\ 
	& $\left( 0.5,N\right) $ & $3.068$ & $3.896$ &  & $2.483$ & $3.153$ \\ 
	&  & $\left[ 1.976,4.542\right] $ & $\left[ 2.731,5.241\right] $ &  & $\left[
	1.602,3.504\right] $ & $\left[ 2.302,4.174\right] $ \\ 
	& $\left( 0.2,L\right) $ & $0.638$ & $0.754$ &  & $0.315$ & $0.434$ \\ 
	&  & $\left[ 0.358,1.060\right] $ & $\left[ 0.494,1.250\right] $ &  & $\left[
	0.190,0.515\right] $ & $\left[ 0.272,0.650\right] $ \\ 
	& $\left( 0.5,L\right) $ & $1.413$ & $1.310$ &  & $0.667$ & $0.707$ \\ 
	&  & $\left[ 0.728,2.472\right] $ & $\left[ 0.763,2.112\right] $ &  & $\left[
	0.381,1.199\right] $ & $\left[ 0.439,1.114\right] $ \\ \hline
\end{tabular}
		\caption{Median, and first and third quartiles [$Q_1,Q_3$], of $100 \times ISE$ for GSS and NP (nonparametric) deconvolution estimators with the bandwidth optimally chosen, $NSR = \sigma_U^2/\sigma_X^2$, measurement error $U$ is $N=\mathrm{Normal}$ and $L=\mathrm{Laplace}$.}
		\label{tab:ISE_opt_estims}
	\end{center}
\end{table}

\begin{table}[h!]
	\begin{center}
		\begin{tabular}{ccccccc}
			\cline{3-7}
			&  & \multicolumn{5}{c}{MISE bandwidth and solution selection $\left(
				n=500\right) $} \\ \hline
			$\pi $ & $\left( \mathrm{NSR},U\right) $ & $\mathrm{MIN}$ & $\mathrm{SKW}$ & 
			\textrm{PHS} & $\mathrm{RND}$ & $\mathrm{NP}$ \\ \hline
			$\pi _{0}$ & $\left( 0.2,N\right) $ & 0.143 & 0.234 & 0.180 & 0.215 & 0.334
			\\ 
			&  & $\left[ 0.069,0.280\right] $ & $\left[ 0.119,0.411\right] $ & $\left[
			0.085,0.332\right] $ & $\left[ 0.107,0.391\right] $ & $\left[ 0.214,0.480%
			\right] $ \\ 
			& $\left( 0.5,N\right) $ & 0.320 & 0.503 & 0.382 & 0.529 & 0.728 \\ 
			&  & $\left[ 0.132,0.671\right] $ & $\left[ 0.259,0.970\right] $ & $\left[
			0.145,0.845\right] $ & $\left[ 0.247,1.063\right] $ & $\left[ 0.488,1.035%
			\right] $ \\ 
			& $\left( 0.2,L\right) $ & 0.173 & 0.196 & 0.202 & 0.216 & 0.233 \\ 
			&  & $\left[ 0.084,0.315\right] $ & $\left[ 0.093,0.372\right] $ & $\left[
			0.097,0.384\right] $ & $\left[ 0.100,0.403\right] $ & $\left[ 0.140,0.395%
			\right] $ \\
			& $\left( 0.5,L\right) $ & 0.286 & 0.317 & 0.350 & 0.384 & 0.401 \\ 
			&  & $\left[ 0.130,0.649\right] $ & $\left[ 0.144,0.776\right] $ & $\left[
			0.162,0.805\right] $ & $\left[ 0.178,0.823\right] $ & $\left[ 0.220,0.674%
			\right] $ \\ \hline
			$\pi _{1}$ & $\left( 0.2,N\right) $ & 1.545 & 1.832 & 1.788 & 1.812 & 2.064
			\\ 
			&  & $\left[ 1.246,1.953\right] $ & $\left[ 1.418,2.289\right] $ & $\left[
			1.467,2.156\right] $ & $\left[ 1.447,2.246\right] $ & $\left[ 1.626,2.560%
			\right] $ \\
			& $\left( 0.5,N\right) $ & 2.474 & 3.166 & 2.640 & 3.052 & 3.810 \\ 
			&  & $\left[ 1.892,3.189\right] $ & $\left[ 2.309,3.974\right] $ & $\left[
			2.011,3.974\right] $ & $\left[ 2.218,4.078\right] $ & $\left[ 3.070,4.705%
			\right] $ \\ 
			& $\left( 0.2,L\right) $ & 0.671 & 1.016 & 0.784 & 1.234 & 1.271 \\ 
			&  & $\left[ 0.402,1.123\right] $ & $\left[ 0.531,1.638\right] $ & $\left[
			0.438,1.310\right] $ & $\left[ 0.646,1.714\right] $ & $\left[ 0.942,1.694%
			\right] $ \\
			& $\left( 0.5,L\right) $ & 0.992 & 1.431 & 1.039 & 1.671 & 1.929 \\ 
			&  & $\left[ 0.526,1.543\right] $ & $\left[ 0.722,2.060\right] $ & $\left[
			0.534,1.693\right] $ & $\left[ 0.983,2.529\right] $ & $\left[ 1.423,2.626%
			\right] $ \\ \hline
			$\pi _{2}$ & $\left( 0.2,N\right) $ & 1.158 & 1.158 & 1.158 & 5.050 & 1.401
			\\ 
			&  & $\left[ 0.809,1.574\right] $ & $\left[ 0.809,1.574\right] $ & $\left[
			0.809,1.574\right] $ & $\left[ 1.179,7.208\right] $ & $\left[ 1.065,1.836%
			\right] $ \\
			& $\left( 0.5,N\right) $ & 3.147 & 3.147 & 3.147 & 6.273 & 3.456 \\ 
			&  & $\left[ 2.025,4.548\right] $ & $\left[ 2.025,4.556\right] $ & $\left[
			2.025,4.594\right] $ & $\left[ 3.174,8.329\right] $ & $\left[ 2.586,4.422%
			\right] $ \\
			& $\left( 0.2,L\right) $ & 0.873 & 0.873 & 0.873 & 4.401 & 0.636 \\ 
			&  & $\left[ 0.537,1.302\right] $ & $\left[ 0.537,1.302\right] $ & $\left[
			0.537,1.302\right] $ & $\left[ 0.881,6.818\right] $ & $\left[ 0.411,0.932%
			\right] $ \\
			& $\left( 0.5,L\right) $ & 1.631 & 1.631 & 1.640 & 3.925 & 1.048 \\ 
			&  & $\left[ 1.031,2.618\right] $ & $\left[ 1.031,2.618\right] $ & $\left[
			1.031,2.653\right] $ & $\left[ 1.557,6.307\right] $ & $\left[ 0.686,1.637%
			\right] $ \\ \hline
		\end{tabular}%
	\caption{Comparison of solution selection methods. Median, as well as first and third quartiles [$Q_1,Q_3$], of $100\times ISE$ for GSS estimator. MIN = choose solution with smallest ISE, SKW = skewness selection method, PHS = phase function selection method, RND = select random ISE value with equal probability, NP = nonparametric deconvolution estimators with plug-in bandwidth, $N=\mathrm{Normal}$, $L=\mathrm{Laplace}$.}
	\label{tab:ISE_MISE500_selection}
	\end{center}
	
\end{table}

\begin{table}[!h]
	\begin{center}
\begin{tabular}{cccccc}
	\\ \hline
	$\pi $ & $\left( \mathrm{NSR},U\right) $ & $\mathrm{CV}$ & $\mathrm{MISE}$ & 
	\textrm{PI} & $\mathrm{NP}$ \\ \hline
	$\pi _{0}$ & $\left( 0.2,N\right) $ & 0.409 & $0.370$ & 0.294 & $0.535$ \\ 
	& $\left( 0.5,N\right) $ & 0.652 & $0.701$ & 0.492 & $1.039$ \\ 
	& $\left( 0.2,L\right) $ & 0.409 & $0.407$ & 0.299 & $0.433$ \\ 
	& $\left( 0.5,L\right) $ & 0.574 & $0.630$ & 0.435 & $0.653$ \\ \hline
	$\pi _{1}$ & $\left( 0.2,N\right) $ & 2.217 & $2.116$ & 2.399 & $2.709$ \\ 
	& $\left( 0.5,N\right) $ & 3.193 & $3.032$ & 3.983 & $4.601$ \\ 
	& $\left( 0.2,L\right) $ & 1.645 & $1.494$ & 1.712 & $1.998$ \\ 
	& $\left( 0.5,L\right) $ & 2.299 & $2.116$ & 2.274 & $2.848$ \\ \hline
	$\pi _{2}$ & $\left( 0.2,N\right) $ & 2.138 & $1.593$ & 1.755 & $1.956$ \\ 
	& $\left( 0.5,N\right) $ & 4.648 & $4.175$ & 3.785 & $4.375$ \\ 
	& $\left( 0.2,L\right) $ & 1.359 & $1.230$ & 0.894 & $1.044$ \\ 
	& $\left( 0.5,L\right) $ & 2.872 & $2.633$ & 2.786 & $1.752$ \\ \hline
\end{tabular}
		\caption{Median of $100 \times \mathrm{ISE}$ for GSS deconvolution estimators with cross-validation (CV), MISE-approximation (MISE) and two-stage plug-in (PI) bandwidths and phase function selection method, as well as nonparametric deconvolution estimator with two-stage plug-in bandwidth ($\mathrm{NP}$). NSR = Noise-to-Signal ratio, measurement error distribution $U$ considered $N=\mathrm{Normal}$, $L=\mathrm{Laplace}$, sample size $n=200$.}
		\label{tab:medianISE_estims200}
	\end{center}
\end{table}

\begin{table}[!h]
	\begin{center}
\begin{tabular}{cccccc}
 \\ \hline
	$\pi $ & $\left( \mathrm{NSR},U\right) $ & $\mathrm{CV}$ & $\mathrm{MISE}$ & 
	\textrm{PI} & $\mathrm{NP}$ \\ \hline
	$\pi _{0}$ & $\left( 0.2,N\right) $ & 0.190 & 0.180 & 0.160 & 0.334 \\ 
	& $\left( 0.5,N\right) $ & 0.356 & 0.382 & 0.297 & 0.728 \\ 
	& $\left( 0.2,L\right) $ & 0.186 & 0.202 & 0.152 & 0.233 \\ 
	& $\left( 0.5,L\right) $ & 0.295 & 0.350 & 0.226 & 0.401 \\ \hline
	$\pi _{1}$ & $\left( 0.2,N\right) $ & 1.885 & 1.788 & 2.027 & 2.064 \\ 
	& $\left( 0.5,N\right) $ & 2.781 & 2.640 & 3.350 & 3.810 \\ 
	& $\left( 0.2,L\right) $ & 0.897 & 0.784 & 0.991 & 1.271 \\ 
	& $\left( 0.5,L\right) $ & 1.264 & 1.039 & 1.304 & 1.929 \\ \hline
	$\pi _{2}$ & $\left( 0.2,N\right) $ & 1.492 & 1.158 & 1.173 & 1.401 \\ 
	& $\left( 0.5,N\right) $ & 3.746 & 3.147 & 2.967 & 3.456 \\ 
	& $\left( 0.2,L\right) $ & 0.845 & 0.873 & 0.471 & 0.636 \\ 
	& $\left( 0.5,L\right) $ & 1.752 & 1.640 & 1.376 & 1.048 \\ \hline
\end{tabular}
		\caption{Median of $100 \times \mathrm{ISE}$ for GSS deconvolution estimators with cross-validation (CV), MISE-approximation (MISE) and two-stage plug-in (PI) bandwidths and phase function selection method, as well as nonparametric deconvolution estimator with two-stage plug-in bandwidth ($\mathrm{NP}$). NSR = Noise-to-Signal ratio, measurement error distribution $U$ considered $N=\mathrm{Normal}$, $L=\mathrm{Laplace}$, sample size $n=500$.}
		\label{tab:medianISE_estims500}
	\end{center}
\end{table}
	
\end{document}